\documentclass[12pt,preprint]{aastex}

\slugcomment{To appear in the 2004 July issue of the Astronomical
Journal.}

\shorttitle{Astrometry of Sources in {\em Spitzer\/} First-Look Survey} 
\shortauthors{Wrobel et al.}
\begin{document}

\title{VLBA Astrometry of Compact Radio Sources \\ 
       in the {\em Spitzer\/} First-Look Survey}

\author{J. M. Wrobel,\altaffilmark{1,2}
        M. A. Garrett,\altaffilmark{3}
        J. J. Condon,\altaffilmark{2,4}
        and R. Morganti\altaffilmark{5}}

\altaffiltext{1}{National Radio Astronomy Observatory, P.O. Box O,
Socorro, NM 87801; jwrobel@nrao.edu}

\altaffiltext{2}{The National Radio Astronomy Observatory (NRAO) is a
facility of the National Science Foundation, operated under
cooperative agreement by Associated Universities, Inc.}

\altaffiltext{3}{Joint Institute for VLBI in Europe, Postbus 2,
NL-7990 AA Dwingeloo, Netherlands; garrett@jive.nl}

\altaffiltext{4}{National Radio Astronomy Observatory, 520 Edgemont
Road, Charlottesville, VA 22903; jcondon@nrao.edu}

\altaffiltext{5}{Netherlands Foundation for Research in Astronomy,
Postbus 2, NL-7990 AA Dwingeloo, Netherlands; morganti@astron.nl}

\begin{abstract}
The NRAO Very Long Baseline Array (VLBA) was used at 1.4~GHz to image
20 compact sources in the inner 1\arcdeg\, radius of the {\em
Spitzer\/} VLA First-Look Survey.  Twelve sources were detected with
the VLBA, with peak flux densities above 2~mJy at 9-mas resolution.
The positions of these VLBA detections have been measured in the
International Celestial Reference Frame (ICRF) to 1$\sigma$ accuracies
of 8~mas East-West and 4~mas North-South.  These positions can help
anchor ancillary data for the First-Look Survey to the ICRF and are
being reported quickly to the community.
\end{abstract}

\keywords{astrometry --- galaxies: active --- radio continuum ---
surveys}

\section{Introduction}

Wide fields observed across the electromagnetic spectrum are beginning
to be surveyed at 10-mas resolution, using the techniques of Very Long
Baseline Interferometry (VLBI) to study sources stronger than about
0.1~mJy at 1.4~GHz \citep{gar01,gar04}.  The corresponding linear
resolution is 90~pc or finer for a {\em WMAP\/} cosmology
\citep{spe03}.  This faint radio population contains sources that are
dominantly energized by star-formation, plus sources -- the active
galaxies -- that are dominantly energized by massive black holes.
VLBI surveys of these faint sources have two primary science goals:
(1) for the star-forming galaxies, seek evidence for embedded
candidate active galactic nuclei or else limit contributions from such
nuclei; and (2) for the active galaxies, seek evidence for the active
nuclei or else limit contributions from such nuclei.  Being unaffected
by obscuration, these VLBI surveys thus help constrain models for the
growth of massive black holes over cosmic time.  Moreover, although
these VLBI surveys must be deep to probe the faint population, their
limiting sensitivities are still such that confusing a luminous type
IIn supernova like SN 1988Z \citep{wil02} for a candidate active
galactic nucleus is a potential problem only for redshifts below
one-tenth.

There are about 700 faint sources per square degree at 1.4~GHz in the
VLA counterpart to the {\em Spitzer\/} First-Look Survey
\citep[FLSVLA;][]{con03}.  That VLA survey had a resolution of
5\arcsec\, FWHM and uniformly covered the circular area with radius
1\arcdeg\ that was guaranteed to be surveyed by {\em Spitzer\/}
(Figure~1).  Given the areal densities of the FLSVLA sources, it
becomes both feasible and efficient to survey the faint radio
population using VLBI phase referencing in the in-beam style
\citep{wro00}.\footnote{Available at
http://www.aoc.nrao.edu/vlba/html/MEMOS/scimemos.html.}  Such VLBI
surveys require substantial time, telescope, and correlator resources,
and therefore must be carefully optimized for maximum scientific
return.  Two key optimizing factors are the suitability of the in-beam
phase calibrators used and the number of faint sources that appear
adjacent to those in-beam phase calibrators.

As a step toward quantifying these factors within the 1\arcdeg\,
radial region shown in Figure~1, the NRAO Very Long Baseline Array
\citep[VLBA;][]{nap94} was used to conduct phase-referenced
observations at 1.4~GHz, in the nodding style \citep{wro00}, of 20
candidate in-beam calibrators that are unresolved and stronger than
5~mJy in the FLSVLA.  Figure~1 shows the locations of these 20 compact
FLSVLA sources, while Table~1 lists for each source the FLSVLA name,
position, position error, and integrated flux density \citep{con03}.
Each compact FLSVLA source has an upper limit to its deconvolved
diameter at FWHM of 2.2\arcsec\, or less, so given the quoted position
errors it is prudent to ensure that the VLBA search region covers a
diameter of about 1-2\arcsec.  Figure~1 also shows the guaranteed
region, of radius 0.25\arcdeg, for the deeper {\em Spitzer\/}
Verification Survey.  Although the rectangular observation regions are
now known for the the First-Look and Verification Surveys, this work
continues to emphasize these circular regions to optimize synergy with
ancillary surveys conducted prior to those {\em Spitzer\/} surveys.
Coincidentally, the region of diameter 0.5\arcdeg\, shown for the
Verification Survey corresponds to the primary beam at FWHM of a VLBA
antenna at 1.4~GHz, a relevant factor in planning in-beam VLBI
surveys.

As the nodding calibrator for the VLBA observations was selected from
the VLBA Calibrator Survey \citep[VCS1;][]{bea02}, all compact FLSVLA
sources detected with the VLBA will have their positions measured in
the International Celestial Reference Frame (Extension 1)
\citep[ICRF-Ext.1;][]{ier99,ma98}.  The positions of these VLBA
detections can, therefore, help anchor ancillary data for the
First-Look Survey to the ICRF, and the purpose of this paper is to
report these positions quickly to the community.  The VLBA
observations and calibration are described in Section~2, while
Section~3 describes the imaging strategies and the astrometric
implications.  A multi-wavelength analysis of both the detections and
non-detections from the VLBA observations will be deferred until more
ancillary data are available for the First-Look Survey.  Also, future
VLBI surveys and multi-wavelength studies will benefit from a new
survey with the Westerbork Synthesis Radio Telescope that complements
the FLSVLA survey by going deeper at 1.4~GHz over a narrower region
with a resolution of 15\arcsec\, FWHM \citep{mor04}.

\section{Observations and Calibration}

The VLBA was used to observe the 20 compact FLSVLA sources and
calibrators on 2004 January 9 UT.  Nine of the 10 VLBA antennas
particpated and provided antenna separations from 240 km to 5800 km.
Data were acquired during about 5 hours in dual circular polarizations
with 4-level sampling and at a center frequency 1.43849~GHz with a
bandwidth of 32~MHz per polarization.  This bandwidth was synthesized
from 4 contiguous baseband channels, each of width 8~MHz.
Phase-referenced observations were made in the nodding style at
elevations above 20\arcdeg.  Successive 80-second observations of
three FLSVLA sources were preceded and followed by a 60-second
observation of the phase, rate, and delay calibrator VCS1 J1722$+$5856
\citep{bea02}, leading to a switching time of 5~minutes.  Switching
angles between that calibrator and a FLSVLA source were 1.8\arcdeg\,
or less (Figure~1).  The assumed position for VCS1 J1722$+$5856 was
$\alpha(J2000.0)=17h22m36.7262s$ and $\delta(J2000.0)=+58\arcdeg
56\arcmin 22.260\arcsec$, with 1$\sigma$ coordinate errors in the
ICRF-Ext.1 of 8~mas and 4~mas, respectively.  Each FLSVLA source was
observed during 6 snapshots spread over time to enhance coverage in
the $(u,v)\/$ plane.  Figure~1 also shows the locations of calibrators
CLASS J1721$+$5926, CLASS J1726$+$6011 \citep{mye03}, and VCS1
J1722$+$6105 \citep{bea02} that were observed similarly to each FLSVLA
source but with switching angles of 0.6, 1.3, and 2.2\arcdeg,
respectively.  Those calibrator observations were intended to assess
phase-referencing conditions; in particular, if such a calibrator can
be detected with an adequate signal-to-noise ratio during an 80-second
observation, then it can be used to quanitify coherence losses.
Finally, a strong calibrator J1740$+$5211 was observed to align the
phases of the independent baseband channels.  Observation and
correlation assumed a coordinate equinox of 2000.

Data editing and calibration were done with the 2004 December 31
release of the NRAO AIPS software, using automatic scripts that follow
the strategies outlined in Appendix C of the NRAO AIPS Cookbook
\footnote{http://www.aoc.nrao.edu/aips/cook.html}.  Diagnostic tables
generated by the scripts were scrutinized to confirm the calibration
quality.  Data deletion was based on system flags recorded at
observation and tape weights recorded at correlation.  Corrections for
the dispersive delays caused by the Earth's ionosphere were made using
electron-content models based on Global Positioning System data and
derived at the Jet Propulsion Laboratory.  VLBA system temperatures
and gains were used to set the amplitude scale to an accuracy of about
5\%, after first correcting for sampler errors.  The visibility data
for the phase, rate, and delay calibrator VCS1 J1722$+$5856 were used
to generate phase-referenced visibility data for each FLSVLA source
and for the calibrators CLASS J1721$+$5926, CLASS J1726$+$6011, and
VCS1 J1722$+$6105.  In contrast, the visibility data for VCS1
J1722$+$5856 and the strong source J1740$+$5211 were self-calibrated.

\section{Imaging Strategies and Astrometric Implications}

The AIPS task IMAGR was used to image the Stokes $I\/$ emission from
each FLSVLA source and each calibrator.  To reduce side-lobe levels
for these multi-snapshot data from 36 antenna separations, the
visibility data were weighted uniformly with robustness 0.5 and a
sensitivity loss of about 10\% relative to natural weighting was
incurred.  A two-stage approach to the imaging was taken.  The first
image, not cleaned, spanned 2048 $\times$ 1.5~mas in each coordinate
and had a typical angular resolution characterized as an elliptical
Gaussian with FWHM dimensions of 12.5~mas by 6.5~mas aligned nearly
North-South (referred to as a resolution of 9~mas hereafter).  The
VLBA search region is well within the field-of-view (FOV) limits set
by time and bandwidth averaging \citep{wro95}.  The most constraining
FOV limit follows from accepting a 10\% drop in the peak amplitude of
a true point source due to averaging over each 8-MHz baseband channel;
the resulting FOV is elliptical with major axis 3600~mas and minor
axis 1900~mas.  Given the array, observation, and imaging parameters,
a detection threshold of 6$\sigma$ $\sim$ 2~mJy was adopted within the
search region, a square of side 1870~mas.  The right ascension and
declination of the peak of that detection was recorded.

A second image was made for each VLBA detection, including the 12 of
20 FLSVLA sources marked with dark triangles in Figure~1 and all
calibrators.  This second image spanned 512 $\times$ 1.5~mas in each
coordinate and was made by shifting the tangent point, as derived from
the right ascension and declination of the peak in the first image, to
the field center.  These shifted images were cleaned in regions
centered on the detections and spanning 9~mas East-West and 15~mas
North-South for the isolated detections but spanning customized
regions for the extended detections.  These shifts were finite for the
phase-referenced and cleaned images (shown in Figures~2 and 3) but
zero by definition for the self-calibrated and cleaned images of VCS1
J1722$+$5856 and J1740$+$5211.

\subsection{VLBA Non-Detections}

Residual errors during phase referencing will degrade the point-source
sensitivity of the VLBA.  To quantify the effect of this loss of
coherence, individual 80-second observations of calibrator VCS1
J1722$+$6105 were phase self-calibrated, imaged, and cleaned.  (The
two CLASS calibrators were too weak for self-calibration with
individual 80-second observations, so they were not used to quantify
coherence losses.)  The peak intensity in the self-calibrated image of
VCS1 J1722$+$6105 was 1.3 times the peak intensity in the
phase-referenced image.  Observations of this calibrator required a
switching angle of 2.2\arcdeg, but switching angles of less than
1.8\arcdeg\, were needed to reach the FLSVLA sources.  Therefore, a
conservative correction for the loss of point-source sensitivity is
about 1.3.  For the VLBA non-detections in Table~1, the quoted
6$\sigma$ upper limits for a point source have been increased by this
factor.

\subsection{VLBA Detections}

The bright components in the phase-referenced images (Figures~2 and 3)
appear to be slightly resolved, but this apparent resolution is
probably mostly artificial and due to residual phase errors.  An
elliptical Gaussian was fit to each of these VLBA images to yield the
positions and integrated flux densities quoted in Table~1.

For the VLBA positions, the calculated error per coordinate is the
quadratic sum of three terms: (a) the 1$\sigma$ error in the
calibrator position; (b) the 1$\sigma$ error in the differential
astrometry; and (c) the 1$\sigma$ error in the Gaussian fit.  Term (c)
was always found to be less than 1~mas.  Term (b) was estimated
empirically to be less than 1.5~mas, by noting that after ionospheric
corrections the visibility phases still wound through 3-4 turns over
the 5-hour observation and then scaling the corresponding 3-4
synthesized beamwidths by the ratio of the switching angles to a
radian.  This estimate is rough but plausible: median astrometric
errors of up to 3~mas have been measured at 1.7~GHz for calibrator
pairs separated by less than 2\arcdeg\, when ionospheric corrections
are not made \citep{cha04}.  For the present observations, the
position of VCS1 J1722$+$6105 in the ICRF was too poorly known to
serve as a useful check on the differential astrometry.  Compared to
terms (c) and (b), term (a) clearly dominates and gives VLBA position
errors in right ascension and declination that always round to 8~mas
and 4~mas, respectively.  The tabulated VLBA positions can therefore
help anchor ancillary data for the First-Look Survey to the ICRF, to a
1$\sigma$ accuracy of 8~mas East-West and a 1$\sigma$ accuracy of
4~mas North-South.  The VLBA position errors represent more than a
factor of ten improvement over the FLSVLA position errors.  Further
improvements to the VLBA astrometry are quite feasible, since the VLBA
position error is dominated by the uncertainty in the calibrator
position, and the prospects of reducing that uncertainty to the 1-mas
level are quite good (and are being actively pursued).

For the VLBA integrated flux density, no corrections were made for
coherence losses and the tabulated error is the quadratic sum of the
5\% scale error and the error in the Gaussian model.  To within their
combined errors, the integrated flux density from the VLBA photometry
is either equal to, or less than, that quoted from the FLSVLA survey.
In the former case, an inference is that the compact FLSVLA sources
have not varied significantly over the 17-35 months separating the VLA
and VLBA observations.  In the latter case, the most probable
inference is the presence of significant source structure on scales
less than a few arcseconds that is too weak and/or too large to be
imaged with multiple VLBA snapshots.

Table~1 lists both VLA and VLBA positions for 12 sources.  Since the
VLBA position errors are much smaller than the quoted VLA position
errors, the difference VLA minus VLBA is taken to be the true VLA
error for these strong sources.  The 1$\sigma$ VLA error derived in
right ascension is 0.013 seconds of time, or about 0.096\arcsec, while
in declination it is 0.08\arcsec.  Thus these differences confirm the
0.1\arcsec\, quoted for all strong sources by \citet{con03}.
Moreover, the mean offsets are $-0.0088\pm0.0039$ seconds of time in
right ascension and $-0.02\pm0.024\arcsec$ in declination, so the VLA
frame is off by a little over 2$\sigma$ in right ascension but is fine
within 1$\sigma$ in declination.  Careful users of the FLSVLA will
therefore want to add 0.0088 seconds of time to all right ascensions
published by \citet{con03}.

\clearpage

\begin{figure}
\epsscale{0.85}
\plotone{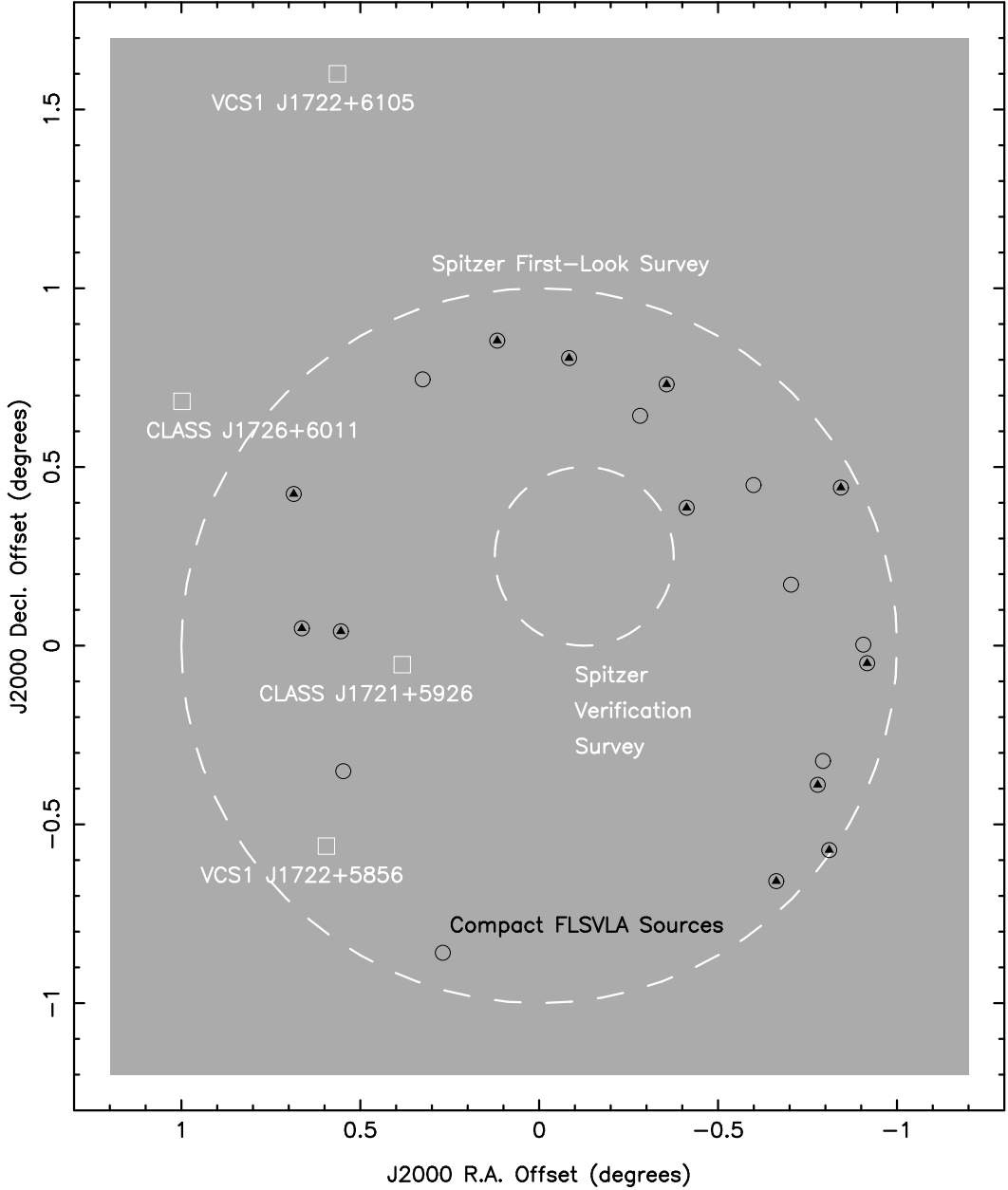}
\caption{Geometry for the VLBA observations.  Solid dark circles show
the locations of compact FLSVLA sources that are unresolved and
stronger than 5~mJy at 1.4~GHz, with inner dark triangles marking the
VLBA detections reported in Table~1.  Light symbols show the locations
of calibrator sources.  Although the {\em Spitzer\/} surveys have
covered rectangular regions on the sky, those regions are guaranteed
to include the areas enclosed by the large dashed circle centered at
$\alpha(J2000.0)=17h18m$ and $\delta(J2000.0)=+59\arcdeg30\arcmin$ for
the First-Look Survey; and by the small dashed circle centered at
$\alpha(J2000.0)=17h17m$ and $\delta(J2000.0)=+59\arcdeg45\arcmin$ for
the deeper Verification Survey.}
\end{figure}

\clearpage

\begin{figure}
\epsscale{0.90}
\plotone{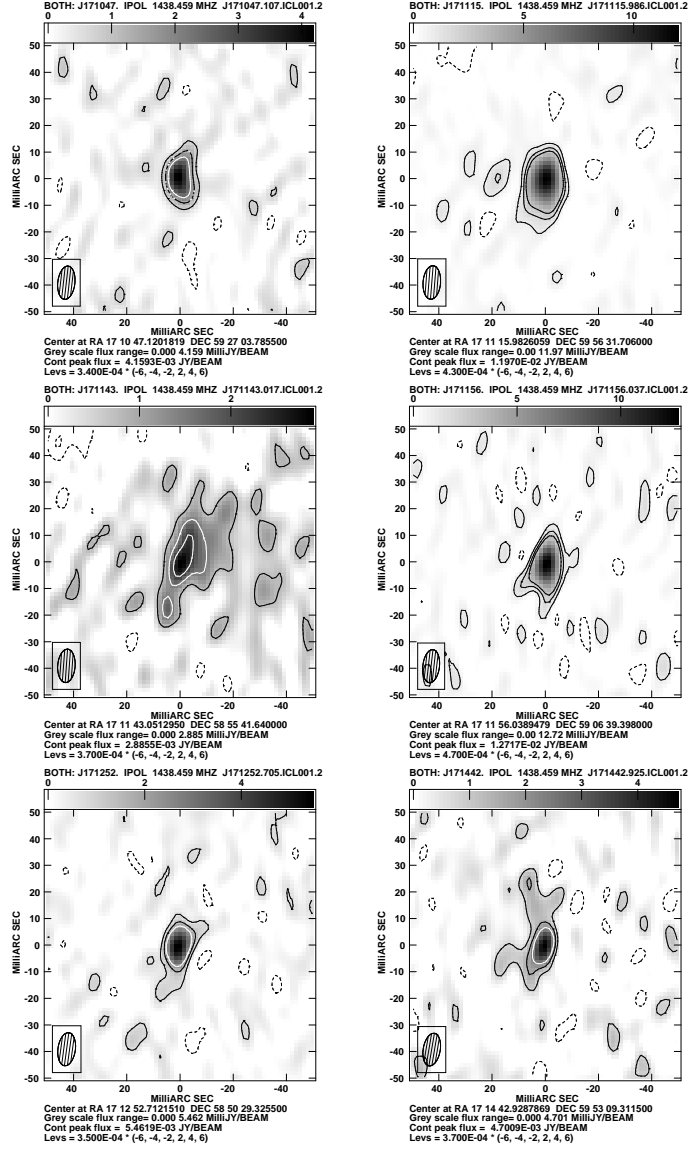}
\caption{Phase-referenced images at 1.43849~GHz of Stokes $I\/$
emission for compact FLSVLA sources detected with the NRAO VLBA on
2004 January 9 UT.  Boxed ellipse shows the Gaussian restoring beam at
FWHM.  Contours are at $\pm$2, $\pm$4, and $\pm$6 times the quoted
1$\sigma$ noise level.  Left to right, starting from the top: FLSVLA
J171047.1$+$592703, J171115.9$+$595631, J171143.0$+$585541,
J171156.0$+$590639, J171252.7$+$585029, J171442.9$+$595309.}
\end{figure}

\clearpage

\begin{figure}
\epsscale{0.90}
\plotone{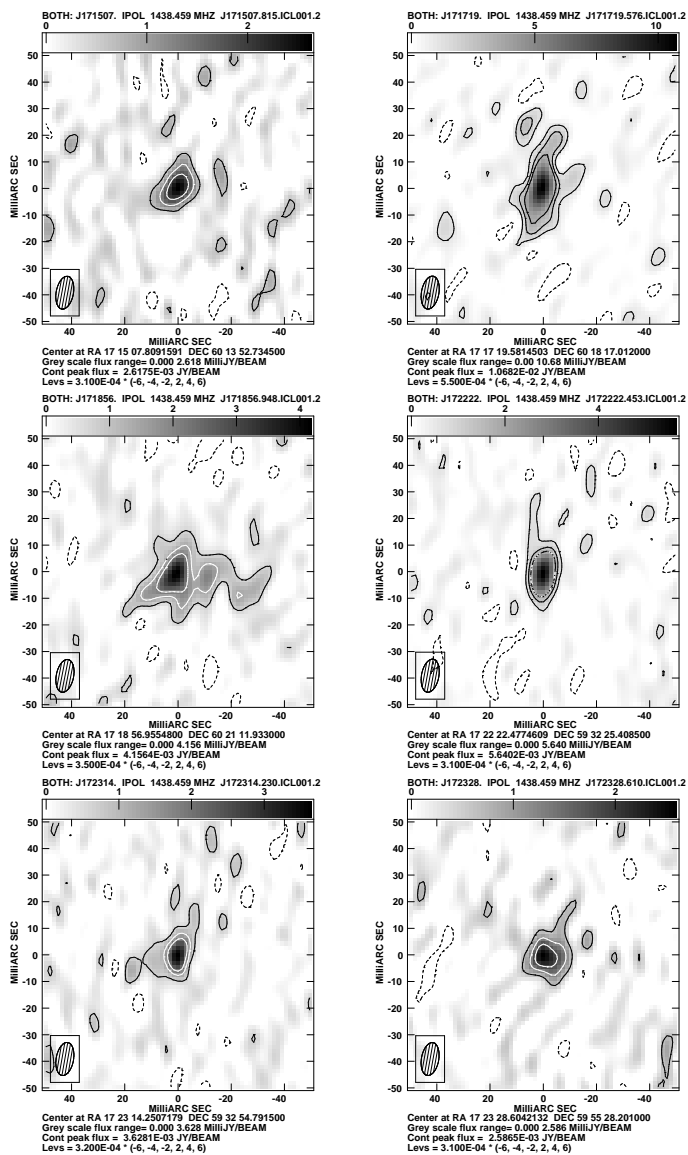}
\caption{As for Figure~2.  Left to right, starting from the top:
FLSVLA J171507.8$+$601352, J171719.5$+$601816, J171856.9$+$602111,
J172222.4$+$593225, J172314.2$+$593254, J172328.6$+$595528.}
\end{figure}

\clearpage

\begin{deluxetable}{lllcccc}
\tabletypesize{\scriptsize} 
\tablecaption{Astrometry and Photometry of Compact 
              FLSVLA Sources at 1.4~GHz\label{tab1}}
\tablewidth{0pc}
\tablehead{ 
\colhead{                 }&
\colhead{ Right Ascension }&
\colhead{ Declination     }&
\colhead{ Position Error  }&
\colhead{ Integrated      }&
\colhead{                 }&
\colhead{                 }\\
\colhead{                 }&
\colhead{ J2000.0         }&
\colhead{ J2000.0         }&
\colhead{ $\sigma_\alpha$,$\sigma_\delta$}&
\colhead{ Flux Density    }&
\colhead{                 }&
\colhead{                 }\\
\colhead{ FLSVLA Name     }&
\colhead{ (h,m,s)         }&
\colhead{ (\arcdeg,\arcmin,\arcsec)}&
\colhead{ (mas,mas)       }&
\colhead{ (mJy)           }&
\colhead{ Ref.            }&
\colhead{ Fig.            }\\
\colhead{ (1)}& \colhead{ (2)}& \colhead{ (3)}& \colhead{ (4)}& 
\colhead{ (5)}& \colhead{ (6)}& \colhead{ (7)}\\
}
\startdata
J171047.1$+$592703& 17 10 47.107 & $+$59 27 03.82 & 100, 100&   7.97$\pm$0.34& 1& ...\\
                  & 17 10 47.1202& $+$59 27 03.786&   8, 4  &   6.57$\pm$0.85& 2&   2\\
J171051.6$+$593010& 17 10 51.682 & $+$59 30 10.96 & 100, 100&   6.79$\pm$0.29& 1& ...\\
                  &           ...&     ...& ...&      $<$2.4\tablenotemark{a}& 2& ...\\
J171115.9$+$595631& 17 11 15.986 & $+$59 56 31.58 & 100, 100&  27.61$\pm$1.17& 1& ...\\
                  & 17 11 15.9826& $+$59 56 31.705&   8, 4  &  20.78$\pm$1.50& 2&   2\\
J171143.0$+$585541& 17 11 43.017 & $+$58 55 41.76 & 100, 100& 113.33$\pm$4.81& 1& ...\\
                  & 17 11 43.0512& $+$58 55 41.640&   8, 4  & 
                                               2.92$\pm$0.40\tablenotemark{a}& 2&   2\\
J171148.5$+$591038& 17 11 48.526 & $+$59 10 38.87 & 100, 100&  35.82$\pm$1.52& 1& ...\\
                  &           ...&     ...& ...&      $<$2.7\tablenotemark{a}& 2& ...\\
J171156.0$+$590639& 17 11 56.037 & $+$59 06 39.26 & 100, 100&  16.48$\pm$0.70& 1& ...\\
                  & 17 11 56.0389& $+$59 06 39.397&   8, 4  &  17.03$\pm$1.29& 2&   2\\
J171225.4$+$594014& 17 12 25.462 & $+$59 40 14.98 & 100, 100&  16.20$\pm$0.69& 1& ...\\
                  &           ...&     ...& ...&      $<$2.6\tablenotemark{a}& 2& ...\\
J171252.7$+$585029& 17 12 52.705 & $+$58 50 29.30 & 100, 100&   7.53$\pm$0.32& 1& ...\\
                  & 17 12 52.7122& $+$58 50 29.325&   8, 4  &   6.81$\pm$0.77& 2&   2\\
J171312.9$+$595657& 17 13 12.936 & $+$59 56 57.99 & 100, 100&  12.52$\pm$0.53& 1& ...\\
                  &           ...&     ...& ...&      $<$2.7\tablenotemark{a}& 2& ...\\
J171442.9$+$595309& 17 14 42.925 & $+$59 53 09.25 & 100, 100&   8.29$\pm$0.35& 1& ...\\
                  & 17 14 42.9290& $+$59 53 09.311&   8, 4  &   6.78$\pm$0.93& 2&   2\\
J171507.8$+$601352& 17 15 07.815 & $+$60 13 52.70 & 100, 100&  10.36$\pm$0.44& 1& ...\\
                  & 17 15 07.8092& $+$60 13 52.735&   8, 4  &   4.61$\pm$0.80& 2&   3\\
J171544.0$+$600835& 17 15 44.021 & $+$60 08 35.40 & 100, 100&  12.61$\pm$0.54& 1& ...\\
                  &           ...&     ...& ...&      $<$2.6\tablenotemark{a}& 2& ...\\
J171719.5$+$601816& 17 17 19.576 & $+$60 18 16.97 & 100, 100&  17.31$\pm$0.74& 1& ...\\
                  & 17 17 19.5816& $+$60 18 17.012&   8, 4  &  19.24$\pm$1.76& 2&   3\\
J171856.9$+$602111& 17 18 56.948 & $+$60 21 11.84 & 100, 100&  14.11$\pm$0.60& 1& ...\\
                  & 17 18 56.9556& $+$60 21 11.932&   8, 4  &  11.07$\pm$1.44& 2&   3\\
J172004.1$+$583827& 17 20 04.153 & $+$58 38 27.62 & 100, 100& 160.81$\pm$6.82& 1& ...\\
                  &           ...&     ...& ...&      $<$2.5\tablenotemark{a}& 2& ...\\
J172037.4$+$601442& 17 20 37.485 & $+$60 14 42.80 & 100, 100&   5.34$\pm$0.23& 1& ...\\
                  &           ...&     ...& ...&      $<$2.4\tablenotemark{a}& 2& ...\\
J172216.1$+$590856& 17 22 16.185 & $+$59 08 56.74 & 100, 100&   6.58$\pm$0.28& 1& ...\\
                  &           ...&     ..& ...&       $<$2.4\tablenotemark{a}& 2& ...\\
J172222.4$+$593225& 17 22 22.453 & $+$59 32 25.47 & 100, 100&   6.96$\pm$0.30& 1& ...\\
                  & 17 22 22.4775& $+$59 32 25.408&   8, 4  &   6.94$\pm$0.74& 2&   3\\
J172314.2$+$593254& 17 23 14.230 & $+$59 32 54.85 & 100, 100&   6.85$\pm$0.29& 1& ...\\
                  & 17 23 14.2508& $+$59 32 54.792&   8, 4  &   5.03$\pm$0.78& 2&   3\\
J172328.6$+$595528& 17 23 28.610 & $+$59 55 28.18 & 100, 100&  12.50$\pm$0.53& 1& ...\\
                  & 17 23 28.6040& $+$59 55 28.201&   8, 4  &   5.49$\pm$0.95& 2&   3\\
\enddata
\tablenotetext{a}{6-$\sigma$ upper limit to the peak flux density at 9-mas resolution.}
\tablerefs{(1) Condon et al. 2003; (2) this work.}
\end{deluxetable}
\end{document}